\documentclass[12pt]{article} 
\usepackage{graphicx,amsmath}
\usepackage{units}
\usepackage{comment}
\usepackage{lineno,color}

\usepackage{multirow}
\usepackage{booktabs}

\newlength{\dinwidth}
\newlength{\dinmargin}
\def\lapproxeq{\lower .7ex\hbox{$\;\stackrel{\textstyle                                                    
<}{\sim}\;$}}                                                    
\def\gapproxeq{\lower .7ex\hbox{$\;\stackrel{\textstyle                                                    
>}{\sim}\;$}}                                                    
\def\be{\begin{equation}}                                                    
\def\ee{\end{equation}}                                                    
\def\bea{\begin{eqnarray}}                                                    
\def\eea{\end{eqnarray}}

\def\sh{\hat s}
\def\sh2{{\hat s}^2}

\begin{document}
                                           
\titlepage                                                    
\begin{flushright}               
IRMP-CP3-26-26 \\
IPPP/26/59  \\                \vspace*{0.5cm}                     
\today \\                                                  
\end{flushright} 
\vspace*{0.5cm}
\begin{center}                                                    
{\Large \bf Low-$x$ parton densities accounting for absorptive effects and  exclusive $J/\psi$ and $\Upsilon$ data from the LHC}\\
\vspace*{1cm}
                                                   
V.~Bertone$^a$, C.~A.~Flett$^b$, A.~D.~Martin$^c$ and M.~G.~Ryskin$^{d}$ \\                                                    
                                                   
\vspace*{0.5cm}                                                    
$^a${\it IRFU, CEA, Université Paris-Saclay, 91191, Gif-sur-Yvette, France}\\                                                   
$^b${\it Centre for Cosmology, Particle Physics and Phenomenology (CP3), Université Catholique de Louvain,
Chemin du Cyclotron, Louvain-la-Neuve, B-1348, Belgium}\\
$^c${\it Institute for Particle Physics Phenomenology, Durham University, Durham, DH1 3LE, U.K.} \\                        
$^d${\it NRC Kurchatov Institute, PNPI, Gatchina, 188300, Russia}

\vspace*{1cm}

\begin{abstract}

  Based on the \textcolor{black}{combined} Deep-Inelastic Scattering (DIS) data \textcolor{black}{from HERA} and the exclusive $J/\psi$ and $\Upsilon$ cross sections measured at the LHC, we investigate the role of absorptive corrections on the low-$x$ behaviour of parton distribution functions~(PDFs) at \textcolor{black}{low and moderate} scales. \textcolor{black}{Using \texttt{xFitter}, we study the impact of these data and of non-linear evolution on PDFs at next-to-leading order~(NLO) and through the NNLO$^*$ method introduced in a previous work. Finally, we determine the effective size of proton hot spots within this framework and discuss the implications of our results for future global PDF determinations.}
  
\end{abstract}
\end{center}  

\vspace*{1cm}

\section{Introduction}
\textcolor{black}{The behaviour of parton distribution functions (PDFs) at low values of the longitudinal momentum fraction $x$ plays a central role in our understanding of high-energy hadronic interactions. Although inclusive deep-inelastic scattering (DIS) measurements from HERA provide the primary constraints on PDFs, the limited kinematic coverage at very small $x$ leads to rapidly increasing uncertainties, particularly at low factorisation scales~\cite{1,2,3}. Exclusive heavy-vector-meson production at the LHC offers a complementary probe of this regime. In particular, the broad rapidity coverage of the Large-Hadron-Collider (LHC) experiments has enabled precise measurements of the differential cross sections for the exclusive production of heavy vector mesons such as $J/\psi$ and $\Upsilon$~\cite{4,5,6,7}. These data probe the gluon PDF down to $x \simeq 3 \times 10^{-6}$ and at
factorisation scales $\mu_F \simeq m_q$, where $m_q$ is the mass of the heavy quark of flavour $q = c$ (charm), $b$ (bottom). They therefore extend our knowledge of the proton structure into a previously inaccessible kinematic region.}

In a recent paper~\cite{geff}, \textcolor{black}{some of us} proposed a method to include these exclusive vector-meson production data into a conventional parton analysis as  ``effective'' values of gluon PDF data points, $xg(x,\mu_F^2)$. As was shown, the inclusion of these new effective gluon data points crucially diminished the low-$x$ PDF uncertainties. However, at such low scales ($\mu_F \simeq m_c$) and at very small $x$, we cannot neglect absorptive effects which will asymptotically lead to the saturation of PDFs~\cite{GLR}. In the present paper, we will, for the first time, consider the role of absorptive corrections in DGLAP evolution, accounting for the latest very low-$x$ vector-meson production data in $pp$ collisions from LHCb experiment. That is, we undertake a combined analysis of inclusive DIS data from HERA and exclusive heavy-vector-meson data from the LHC, in which non-linear (screening) effects are accounted for. 

\textcolor{black}{The paper is organised as follows. In Sect.~2, we introduce the non-linear DGLAP evolution and describe its practical implementation in \texttt{APFEL++}. In Sect.~3, we present the analysis setup implemented in \texttt{xFitter} and discuss the results of fits performed at next-to-leading order~(NLO) using the non-linear evolution framework, combining inclusive DIS data from HERA with effective gluon PDF points extracted from exclusive heavy-vector-meson production measurements at the LHC. In Sect.~\ref{subsec:NNLOstar}, we apply the NNLO* method, introduced in one of our previous works~\cite{geff}, to assess the impact of including exclusive heavy-vector-meson production data in a next-to-next-to-leading order~(NNLO) PDF analysis. In  this approach, DIS data are treated at full NNLO accuracy, while the missing NNLO corrections to the exclusive process are approximated through the determination of a $K$-factor. Finally, we conclude in Sect. 4.}

\section{Steps towards evolution with absorptive corrections}
\textcolor{black}{Since the way we incorporate absorptive corrections into DGLAP evolution is a new procedure, we introduce it step by step.} It is known that when the DGLAP evolution starts from some fixed input PDF, the absorptive corrections diminish the resulting low-$x$ parton densities. The effect is most visible at \textcolor{black}{intermediate scales, neither too high nor too close to the starting scale  $Q_0$ of the evolution~\cite{KMRS,Bar,Esk,Gu}.}

\textcolor{black}{In fitting the data, one must observe the opposite effect. To reproduce the same data when accounting for these absorptive corrections, the input distributions at the starting scale must be larger than those obtained from a fit in which absorptive effects are neglected.}

The first step is to study the pure gluon case. Here the DGLAP $\mu_F^2$ evolution with absorptive corrections included reads~\cite{MQ}
\begin{equation}
\frac{d(xg(x,\mu_F^2))}{d\ln \mu_F^2}=\frac{\alpha_s}{2\pi}\int_x^1 \frac{dx'}{x'}P_{gg}\left(\frac{x}{x'}\right)x'g(x',\mu_F^2)~-~\frac{81\alpha_s^2}{16R^2\mu_F^2}\int\frac{dx'}{x'}(x'g(x',\mu_F^2))^2\ ,
\label{mq}
\end{equation}
where $P_{gg}$ is the gluon–gluon Altarelli–Parisi splitting function, $\alpha_s$ is the QCD coupling at the scale $\mu_F^2$ and the parameter $R$ characterises the transverse spatial extent of the parton distributions in the impact-parameter plane. \textcolor{black}{Typical benchmark values of $R$ are $R=5$ GeV$^{-1}$, corresponding to a homogeneous parton distribution over the whole proton transverse area, and $R=2$ GeV$^{-1}$, corresponding to the so called `hot spot' scenario, in which partons are concentrated in localised regions around the small-size valence quarks.}

Since the leading $\ln(1/x)$ term in $P_{gg}(z)/z$ is
$2N_c/z$, the non-linear correction factor $F$ can be written as
\begin{equation}
F~=~1-\frac{27\alpha_s\pi}{16R^2\mu_F^2}x'g(x',\mu_F^2)\ .
\label{mq2}
\end{equation}
As can be seen in eqn.~(\ref{mq2}), the correction increases as $R$ decreases. 
\textcolor{black}{Moreover, the correction becomes more significant as the gluon density, $x'g(x',\mu_F^2)$, increases in the low-$x$ region.}
   
  Strictly speaking, the factor $(x'g(x',\mu_F^2))^2$ appearing in the last term of eqn.~(\ref{mq}) describes the distribution of {\em two different} gluons and these two gluons may carry different momentum fractions, $x_1$ and $x_2$. That is, one should perform an integration over $x_1,x_2$ under the condition $x_1+x_2=x'$ (see e.g. ~\cite{Z}).
On the other hand, both $x_1$ and $x_2$ are in any case of the order of $x'$ and the effect of this integration may be compensated by variations of the phenomenological parameter $R$.  We will therefore keep the simplified form shown in eqn.~(\ref{mq}).

\begin{figure*}[t]
\begin{center}
\includegraphics[scale=0.25]{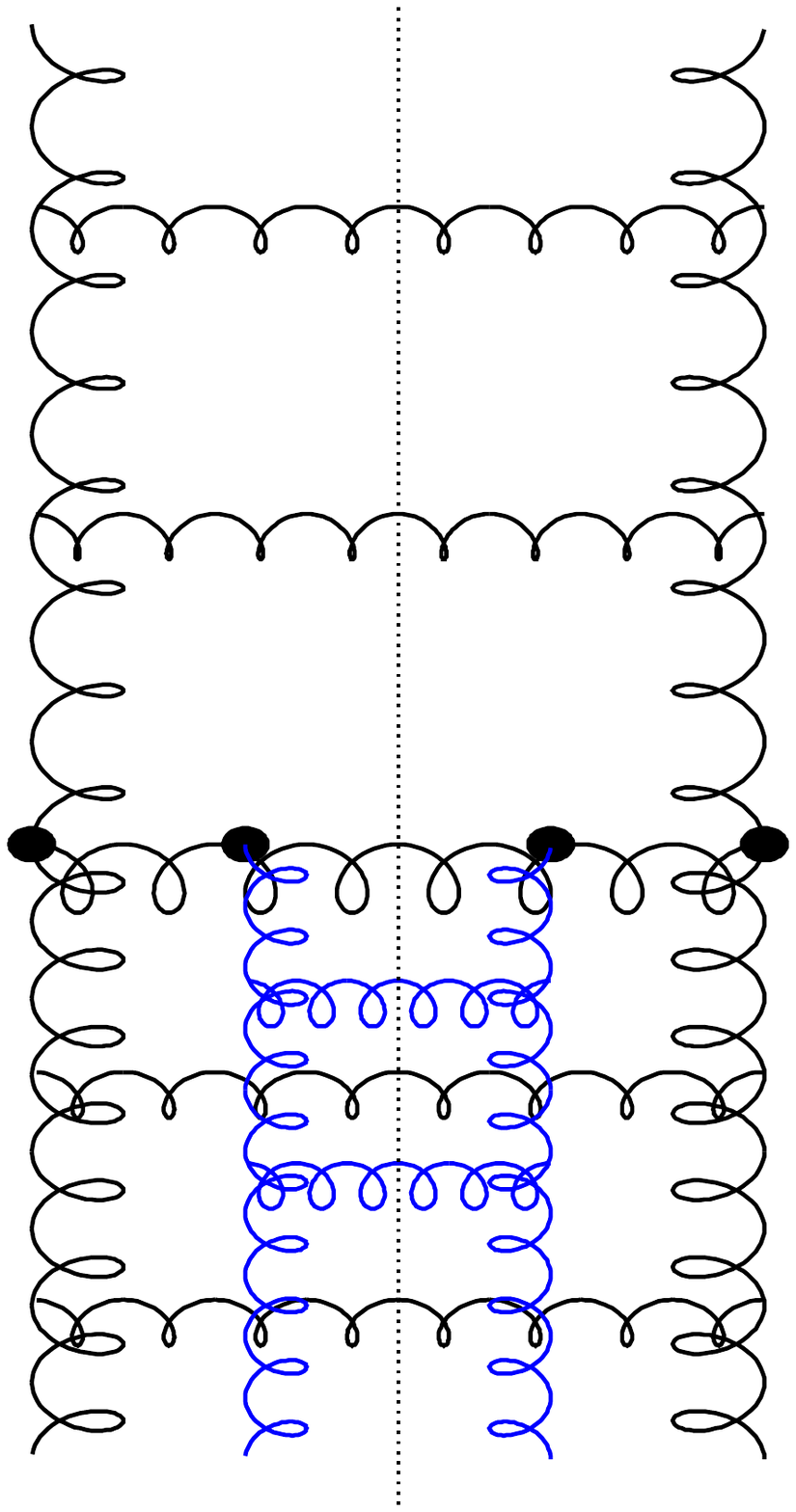}
\hspace{.5cm}
\includegraphics[scale=0.25]{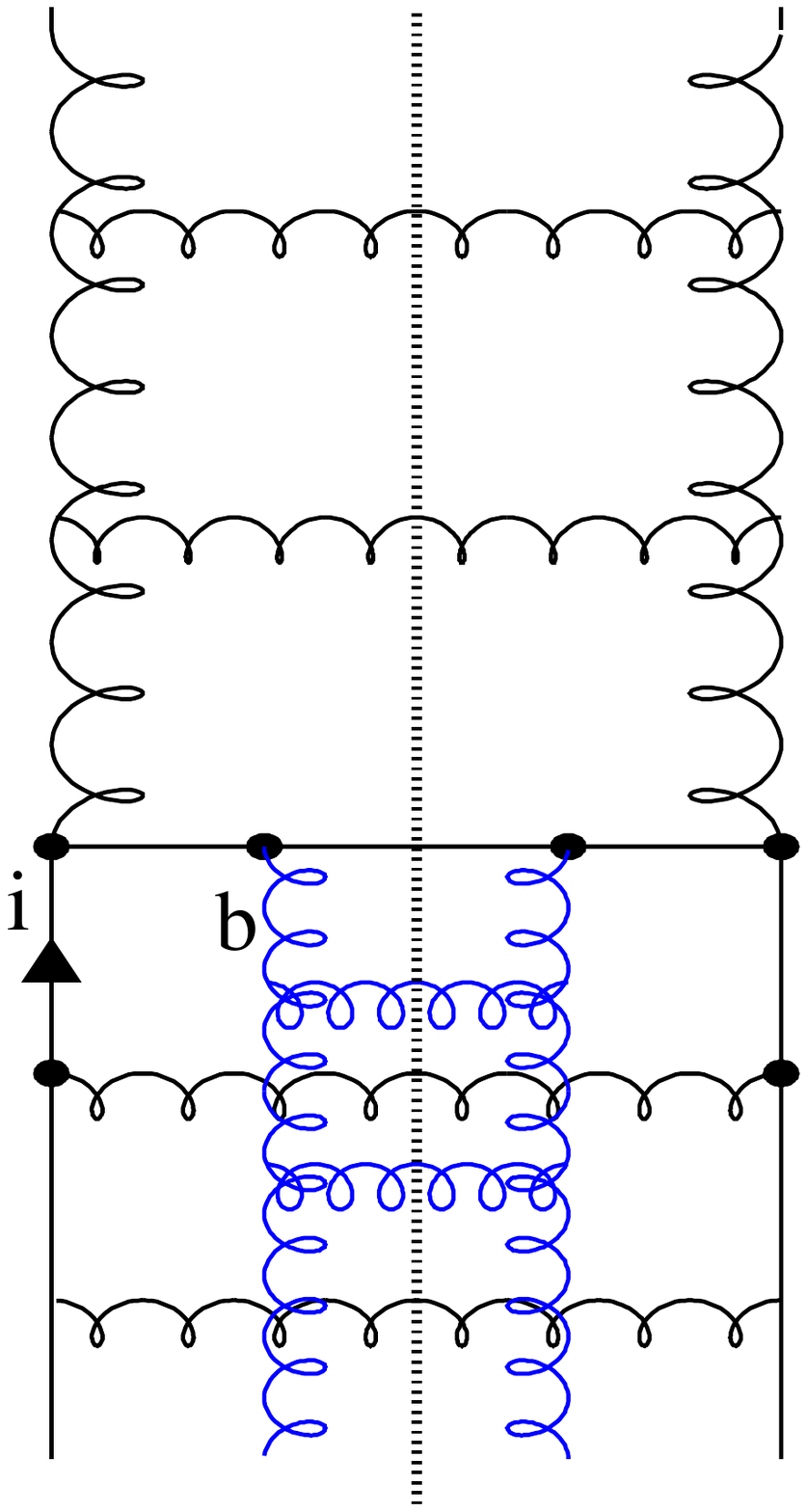}
\caption{\sf Representative diagrams for absorptive corrections during DGLAP evolution, \textcolor{black}{showing an incoming gluon (left) and an incoming quark (right). The additional screening gluons responsible for the non-linear corrections are indicated in blue.} 
}
\label{f1}
\end{center}
\end{figure*}

Since, at low scales and very small $x$ the second term in eqn.~(\ref{mq2}) may exceed unity, \textcolor{black}{we replace it by its eikonalised form by introducing the absorptive screening factor}
\begin{equation}
F_g~
=~\exp\left(-\frac{C_A9\alpha_s\pi}{16R^2\mu_F^2}x'g(x',\mu_F^2)\right),\ 
\label{ab5}  
\end{equation}
\textcolor{black}{which resums the contributions from multiple screening by gluon ladders, similar to that shown in blue in Fig.~\ref{f1}.\footnote{Equation~\eqref{ab5} is analogous to the Golec-Biernat-Wusthoff model~\cite{GW}.}}

This, together with the inclusion of quarks in the evolution, gives the following two coupled-evolution equations, one for gluons and one for singlet quarks:

\begin{equation}
\frac{d(xg(x,\mu_F^2))}{d\ln \mu_F^2}=\frac{\alpha_s}{2\pi}\int_x^1 \frac{dx'}{x'}\left(P_{gg}\left(\frac{x}{x'}\right)x'g(x',\mu_F^2)\cdot F_g + P_{gq}\left(\frac{x}{x'}\right)x'q(x',\mu_F^2)\cdot F_q\right),
\label{ab3}
\end{equation}
and
\begin{equation}
\frac{d(xq(x,\mu_F^2))}{d\ln \mu_F^2}=\frac{\alpha_s}{2\pi}\int_x^1 \frac{dx'}{x'}\left(P_{qg}\left(\frac{x}{x'}\right)x'g(x',\mu_F^2)\cdot F_g + P_{qq}\left(\frac{x}{x'}\right)x'q(x',\mu_F^2)\cdot F_q\right) 
\label{ab4}\ .
\end{equation}
Here  
\begin{equation}
F_q~
=~\exp\left(-\frac{C_F9\alpha_s\pi}{16R^2\mu_F^2}x'g(x',\mu_F^2)\right),
\label{ab6}  
\end{equation}
is the analogous screening factor for a DGLAP kernel with an initial-state quark, whose colour charge $C_F=(N_c^2-1)/(2N_c)=4/3$ is smaller than the gluon colour charge $C_A=N_c=3$.

\begin{figure*}[t]
\begin{center}
\hspace{-.3cm}
\includegraphics[scale=0.24]{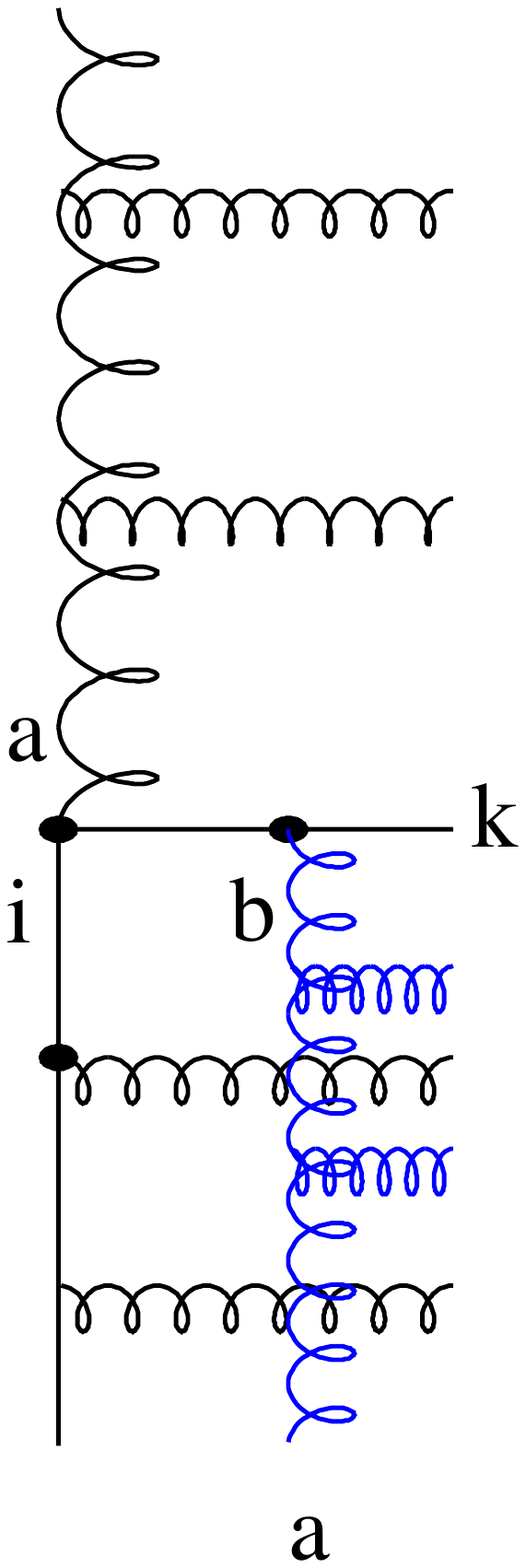}
\hspace{.5cm}
\includegraphics[scale=0.24]{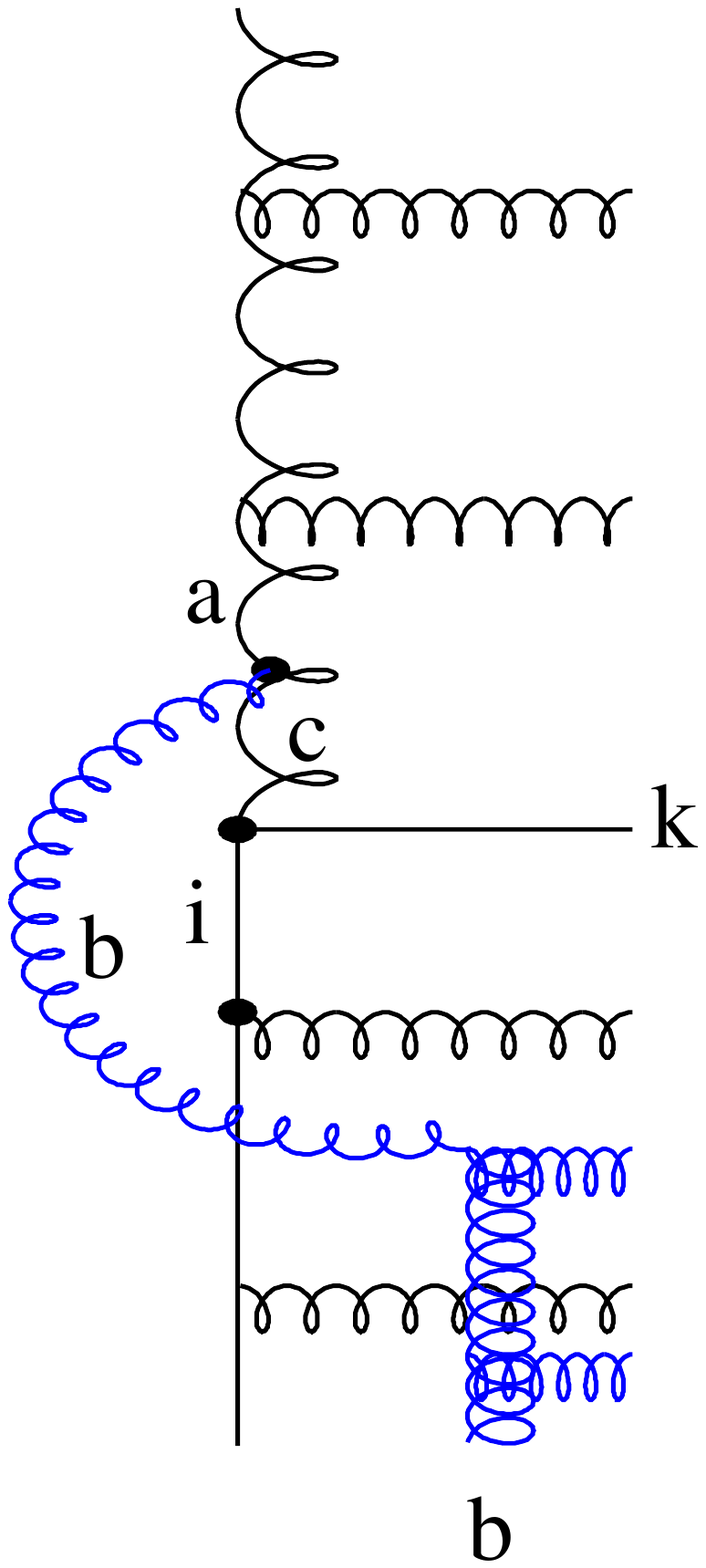}
\hspace{.7cm}
\includegraphics[scale=0.24]{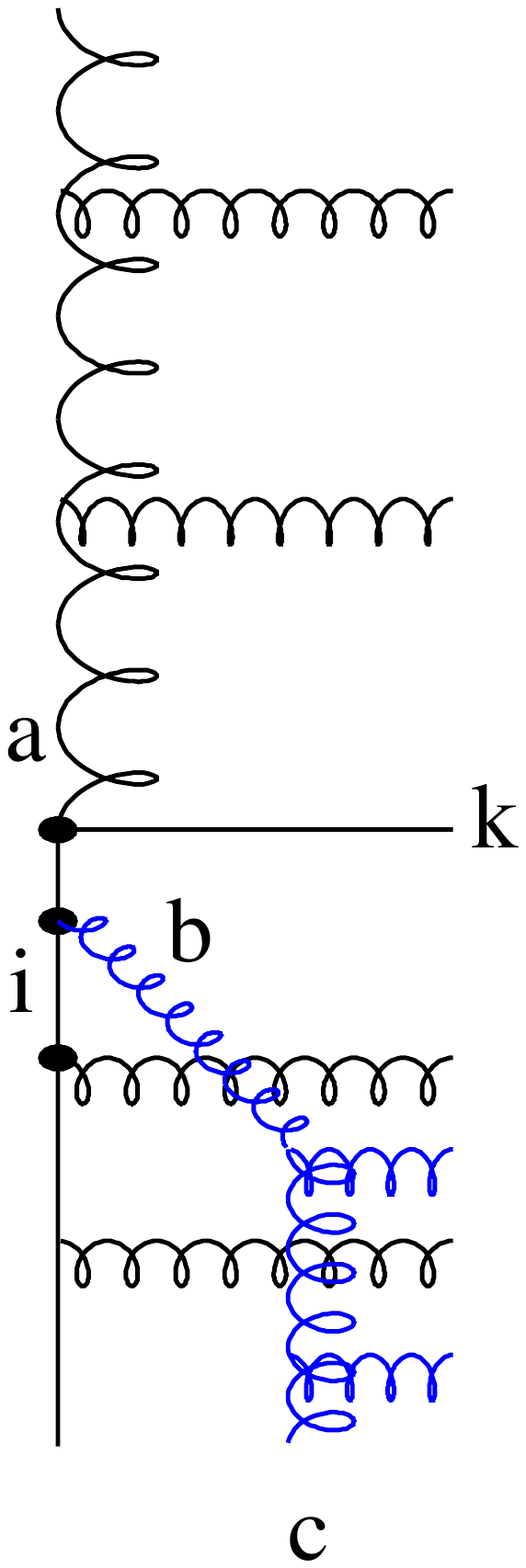}
\caption{\sf \textcolor{black}{Absorptive correction amplitudes. The additional gluon couples to the final $s$-channel parton in (a) and to a $t$-channel parton in (b). Their coherent sum gives the colour factor corresponding to the effective diagram shown in (c).}
}
\label{f2}
\end{center}
\end{figure*}

\begin{figure*}[t]
\begin{center}
\includegraphics[scale=0.25]{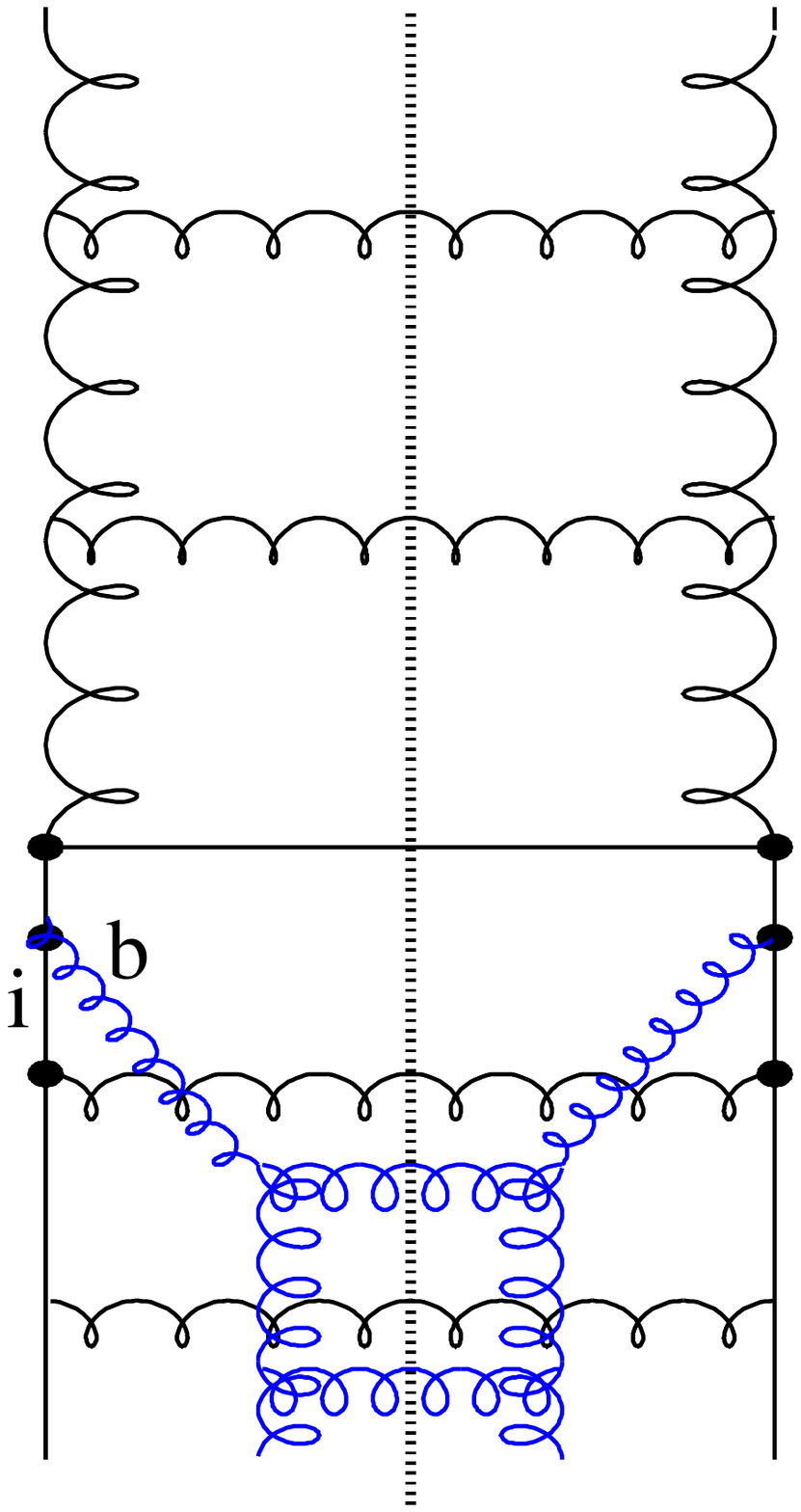}
\caption{\sf \textcolor{black}{Effective diagram representing the absorptive correction colour factor at the cross-section level, obtained after summing over all gluon attachments in the amplitude and complex-conjugate amplitude.}
}
\label{f3}
\end{center}
\end{figure*}

\textcolor{black}{The colour factor appearing in $F_{g,q}$ depends solely on the nature of the {\em incoming} parton. This can be seen explicitly from the representative diagrams shown in Figs.~\ref{f1} and~\ref{f2}. The coherent sum of the amplitudes in which the additional central (blue) gluon interacts with the final-state $s$-channel parton (Fig.~\ref{f2}a) and the outgoing parton in the $t$-channel evolution (Fig.~\ref{f2}b) gives a colour structure equivalent to that obtained when the gluon couples directly to the incoming parton.\footnote{\textcolor{black}{This follows from the identity $-i\bar q_i t^a t^b q_k  - f^{abc} \bar q_i t^c q_k = -i\bar q_i t^b t^a q_k$, where the first term on the left-hand side corresponds to the colour structure of Fig.~\ref{f2}a,  the second term corresponds to that of Fig.~\ref{f2}b, and the right-hand side is precisely the colour structure of Fig.~\ref{f2}c. Here, $i,k$ denote quark colour indices, while $a,b,c$ are gluon colour indices. }} The representative diagrams for the absorptive corrections generated by this additional gluon at the cross-section level are shown in Fig.~\ref{f1}. To illustrate the origin of the colour factor in more detail, let us first consider only the amplitude corresponding to the left-hand side of the cut (i.e. the region to the left of the vertical dotted line) in Fig.~\ref{f1}b. The additional gluon, carrying colour index $b$, can couple either to the final-state quark (Fig.~\ref{f2}a) or to the emitted gluon in the evolution chain (Fig.~\ref{f2}b). The coherent sum of these two contributions gives the colour coefficient corresponding to Fig.~\ref{f2}c, namely the colour structure obtained by attaching the gluon to the incoming parton. Applying the same argument to the complex-conjugate amplitude on the opposite side of the cut, the colour structure of the full cross section is therefore equivalent to that shown in Fig.~\ref{f3}.}

\textcolor{black}{In other words, the value of the absorptive correction $F_{g,q}$ is determined by the colour of the  incoming parton. More precise kinematics for the parton recombination terms in the evolution equations have been implemented in~\cite{Gu} following the approach of~\cite{Z}. In contrast, the simplified treatment adopted here has the advantage that it can be consistently combined with NLO and NNLO evolution kernels.
}

\textcolor{black}{Note that, since the factor $F_{g,q}$ depends on the colour and PDF of the incoming parton, it does not violate the $z\to 1-z$ symmetry ($z=x/x'$) of the splitting functions and therefore preserves all relevant conservation laws.}

The non-linear DGLAP eqns.~(\ref{ab3}) and~(\ref{ab4}) have been
implemented in \texttt{APFEL++}~\cite{Bertone:2013vaa,
  Bertone:2017gds} by extending its native $x$-space evolution
machinery. In the linear case, \texttt{APFEL++} represents the PDFs
and the splitting functions on a discrete interpolation grid in $x$
and performs the integration in $t = \ln \mu_F^2$ by means of a
Runge--Kutta algorithm. The evolved distributions can then be obtained
either by direct integration of the DGLAP equations or, equivalently,
by acting on the input PDFs with pre-computed evolution operators. The
non-linear terms in eqns.~(\ref{ab3})--(\ref{ab4}) involve the
absorptive factors $F_g$ and $F_q$, which depend on the gluon density
at the running scale $Q^2$ and therefore preclude the construction of
$\mu_F^2$-independent evolution operators. We instead solve the coupled
non-linear system directly: at each Runge--Kutta step, $F_g$ and $F_q$
are evaluated from the current $x'g(x',\mu_F^2)$ on the interpolation
grid, the modified kernels are convolved with the PDFs in the standard
way, and the result is used to advance the distributions by one step
in $t$. Because the absorptive factors enter only as multiplicative
modifications of the splitting kernels at the convolution level, the
implementation is fully compatible with the NLO and NNLO splitting
functions already available in \texttt{APFEL++} and leaves the flavour
structure of the linear evolution untouched. This extension has been
interfaced with \texttt{xFitter}~\cite{xF}, so that the parameter $R$
controlling the strength of the absorptive corrections in
eqns.~(\ref{ab5})--(\ref{ab6}) can be determined alongside the input PDF parameters.
 
 \section{Discussion of the results}

 As in our previous~\cite{geff}, we use \texttt{xFitter}~\cite{xF} as a means to
 assess the impact of our new effective data points on PDF
 extractions. As mentioned in Sec. 2, this time non-linear DGLAP
 evolution is provided by an extended version of
 \texttt{APFEL++}~\cite{Bertone:2013vaa, Bertone:2017gds}, which we
 seamlessly integrate with \texttt{xFitter} to allow PDF fits with
 absorptive (non-linear) DGLAP corrections included.

Explicitly, we use \texttt{xFitter} DIS neutral-current~(NC) and charged-current~(CC) data from RunI+II HERA~\cite{H1:2015ubc}, and our effective gluon PDF pseudo data obtained in~\cite{geff} from $J/\psi$ and $\Upsilon$ heavy-quarkonium production data from LHCb. We use the default \texttt{HERAPDF} parametrisation class in \texttt{xFitter} for all parton flavours and the PDF basis \texttt{UvDvUbarDbarS}. That is, the PDFs at the input 
scale $Q_0$ in this basis are of the form 
\begin{align}
    xf_g(x,Q_0) &= A_g x^{B_g} (1-x)^{Cg}, \nonumber \\
    xf_{u_v}(x,Q_0) &=  A_{u_v} x^{B_{u_v}} (1-x)^{C_{u_v}} (1 + E_{u_v} x^2), \nonumber \\
    xf_{d_v}(x,Q_0) &=  A_{d_v} x^{B_{d_v}} (1-x)^{C_{d_v}}, \nonumber \\
    xf_{\bar{D}}(x,Q_0) &=  A_{\bar{D}} x^{B_{\bar{D}}} (1-x)^{C_{\bar{D}}}, \nonumber \\
    xf_{\bar{U}}(x,Q_0) &=  A_{\bar{U}} x^{B_{\bar{D}}} (1-x)^{C_{\bar{U}}} (1 + D_{\bar{U}} x),
\end{align}
where $g, u_v, d_v$ denote the gluon, up and down valence quarks, $\bar{U} = 2\bar{u} = u - u_v$, $\bar{D} = \bar{d} + \bar{s}$, $\bar{d} = (1-f_s)\bar{D}$ and $s=\bar{s} = f_s \bar{D},$ with $f_s = 0.4$. \textcolor{black}{In particular, this implies that we do not use the default \texttt{NegativeGluon} class in \texttt{xFitter}, i.e. we do not assume an input gluon PDF of the form
\begin{equation}
  xf_g(x,Q_0) = A_g x^{B_g} (1-x)^{Cg}  -A'_{g} x^{B'_{g}}(1-x)^{C'_{g}}.
  \label{secondt}
\end{equation} 
This form was introduced} to reproduce the (negative) gluon PDF behaviour obtained in the HERAPDF2.0 global PDF analysis~\cite{H1:2015ubc} at relatively small $x$. \textcolor{black}{The second term, which we hereafter refer to as the `negative gluon component', has $C'_{g} = 25$ fixed in order to suppress its contribution at larger $x$.} \textcolor{black}{As we will show, this term also allows for a partial compensation of the effect of absorptive corrections through a modification of the input distribution}. In our earlier analysis describing DIS data with linear DGLAP evolution~\cite{geff}, this term was included in the gluon input parametrisation to account for possible modifications of the gluon behaviour at {\it very} small $x$. Now, as will be demonstrated, by accounting for absorptive (non-linear)
corrections, we obtain a sufficiently good and better motivated description without the need for this additional term. Here, $Q_0 = 1.54~\text{GeV}$ and the values of the heavy-quark masses set to $m_c = 1.55~\text{GeV}$ and $m_b = 4.73~\text{GeV}$.\footnote{This choice is consistent with the determination of the effective gluon PDF data points at the ``optimal'' factorisation scale choice $\mu_F = m_q$ in our approach at NLO, see~\cite{opt,opt2,Q0}, where we took $m_{c,b}$ at these numerical values.}
The values of $A_{u_v}$ and $A_{d_v}$ are fixed by the valence sum rules while $A_g$ is fixed by the momentum sum rule, and the other parameters $\left\{B_g, C_g, B_{u_v},C_{u_v},E_{u_v},B_{d_v},C_{d_v},A_{\bar{D}},B_{\bar{D}},C_{\bar{D}}, A_{\bar{U}}, C_{\bar{U}},D_{\bar{U}}\right\}$ are fitted.  The DIS data span a wide range in $Q^2$ and are fitted in the FONLL general-mass variable-flavour-number scheme with a kinematic cut at $Q^2 > 2.4~\text{GeV}^2$. Our $J/\psi$ and $\Upsilon$ data points sit at a fixed scale of $Q^2 = m_c^2\simeq 2.4~\text{GeV}^2$ and $Q^2=m_b^2 \simeq 22.4~\text{GeV}^2$, respectively. 

 We now present the \texttt{xFitter} NLO analysis of DIS data supplemented by the effective low-$x$ gluon points obtained from the exclusive  $J/\psi$ and $\Upsilon$ production data from LHC in \cite{geff}. \textcolor{black}{The data included in our fits are listed in Table~\ref{tab:1}.}

 \begin{table}[ht]
\centering
\begin{tabular}[t]{lccc}
\toprule
Dataset&No. of data points\\
\midrule
HERA1+2 NCep 820&73\\
HERA1+2 NCep 460&207\\
HERA1+2 CCep&39\\
HERA1+2 NCem&159\\
HERA1+2 CCem&42\\
HERA1+2 NCep 575&257\\
HERA1+2 NCep 920&391\\
LHC excl. $J/\psi$ $pp$ 13 TeV&10 \\
LHC excl. $\Upsilon$ $pp$ 7,8 TeV&3 \\
\midrule
Total: &1181\\
\bottomrule
\end{tabular}
\caption{\sf{\textcolor{black}{Experimental DIS data points from HERA and exclusive heavy vector meson effective gluon points extracted from LHC data used in the NLO fits.}}} 
\label{tab:1}
\end{table}

 Note that we include the 13~TeV $J/\psi$ dataset in the fit, but not the earlier 7 TeV data set. 
This is because, at the time the 7 TeV data were collected, the experimental apparatus \textcolor{black}{did not permit} a veto on backward particles, so the selected event sample was not purely exclusive.\footnote{We thank Ronan McNulty for discussions concerning the experimental setup via private communication.} 
We compare NLO PDFs with non-linear corrections turned on and off in the \texttt{APFEL++} evolution code. In the case of fitting with non-linear corrections, we additionally determine the parameter $R$ in eqn.~(1) that governs the size of the absorptive corrections by studying the $\chi^2$ figure of merit as a function of $R$. \textcolor{black}{Specifically, we run fits with different values $R$ and the best-fit value is found from the minimum of the $\chi^2$ profile.}

\begin{figure*}[t]
\begin{center}
\hspace{-.5cm}
\includegraphics[scale=0.8]{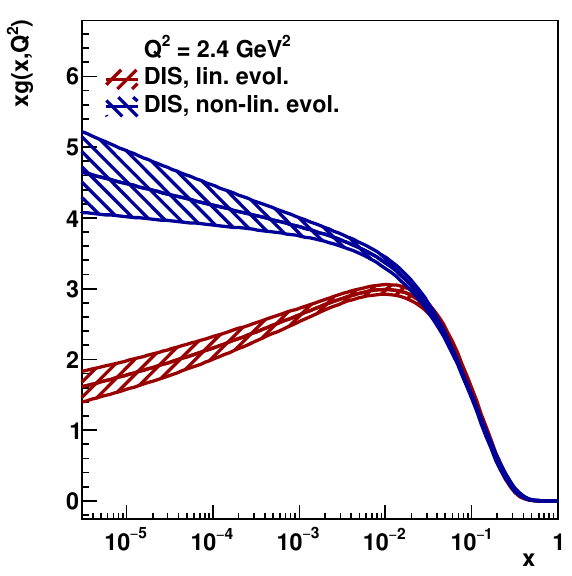}
\qquad
\includegraphics[scale=0.8]{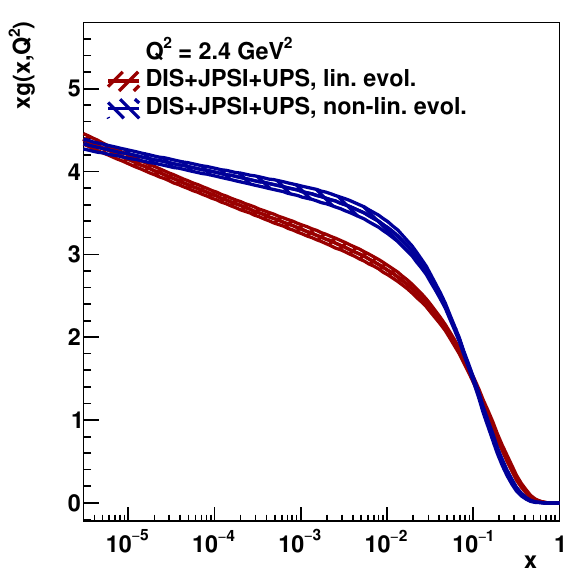}
\caption{\sf The gluon distributions at scale $Q^2=2.4$ GeV$^2$
 obtained {\em with} (blue) and {\em without} (red) absorptive corrections, fitting just DIS data (left) and DIS data combined with exclusive $J/\psi$ and $\Upsilon$ data (right).
 The absorptive corrections were calculated for $R = 3.1$ GeV$^{-1}$ for the fit with just the DIS data and $R=3.6$ GeV$^{-1}$ for the combined fit.
}
\label{f21}
\end{center}
\end{figure*}

 In Fig.~\ref{f21} (left), we show the gluon PDFs obtained from fits to DIS data alone, comparing linear and non-linear evolutions, while in Fig.~\ref{f21} (right), we perform a combined fit of DIS and heavy-quarkonium data, again comparing linear and non-linear evolutions.  It is shown that, in both cases, the inclusion of absorptive corrections generally leads to a larger low-$x$ input gluon distribution \textcolor{black}{in the interval $10^{-5} < x < 0.1$}. \textcolor{black}{At $x \gapproxeq 0.1$, nonlinear effects are negligible, and the smaller gluon density shown in blue in Fig.~\ref{f21} (right) in this region is fixed by the momentum sum rule. At $x \lapproxeq 5 \times 10^{-6}$, however, the exclusive data are too sparse to constrain the small-$x$ power of the gluon PDF. Since the linear-evolution DIS fit yields a smaller gluon density than the non-linear evolution DIS fit of Fig.~\ref{f21} (left) for $x \lapproxeq 10^{-3}$, the inclusion of the exclusive $J/\psi$ and $\Upsilon$ production data,\footnote{As was described in~\cite{geff}, this was done using the `effective' gluon densities $g_{{\rm eff}}(x,Q^2)$.} forces the linear fit towards  an artificially large $\lambda$, with $xg \propto x^{-\lambda}$, in order to reproduce the effective $J/\psi$ data. This, in turn, worsens the description of the DIS data.} Inclusion of the exclusive data essentially diminishes the uncertainties of the obtained results at very low $x$. \textcolor{black}{Tables~\ref{tab:2} and~\ref{tab:3} of Appendix A summarise the partial $\chi^2$ per degree of freedom for each dataset considered.} \textcolor{black}{It can be seen that the DIS data alone are described comparably well by both linear and non-linear evolution, with a modest improvement from $\chi^2$/d.o.f. = 1.20 to 1.18 when non-linear effects are included. Upon adding the exclusive heavy-vector-meson data, the difference becomes more pronounced, with the global $\chi^2$/d.o.f. increasing to 1.25 for linear evolution as indicated above, while remaining essentially unchanged at 1.18 for the non-linear case. The additional exclusive data are themselves described rather well, with $\chi^2$/d.o.f. = 12.6/10 and 2.9/3 for $J/\psi$ and $\Upsilon$, respectively.  This stability of the non-linear fit upon including the exclusive data, together with its lower global $\chi^2$/d.o.f., suggests that the inclusion of absorptive effects provides a more consistent description of the combined DIS and exclusive quarkonium data. }

\begin{figure*}[t]
\begin{center}
\hspace{-.5cm}
\includegraphics[scale=0.8]{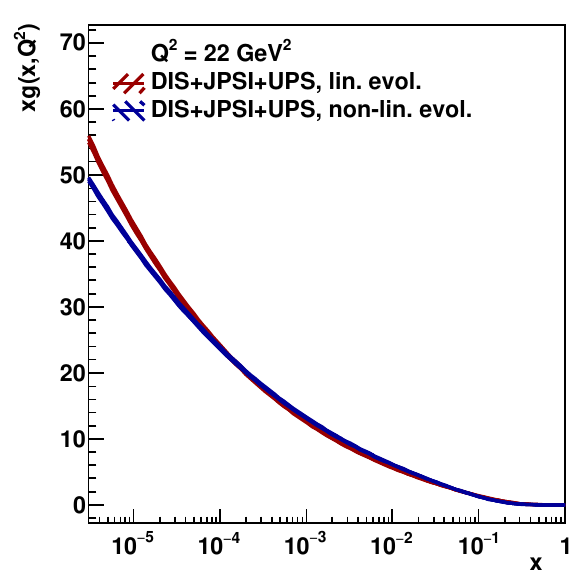}
\qquad
\includegraphics[scale=0.8]{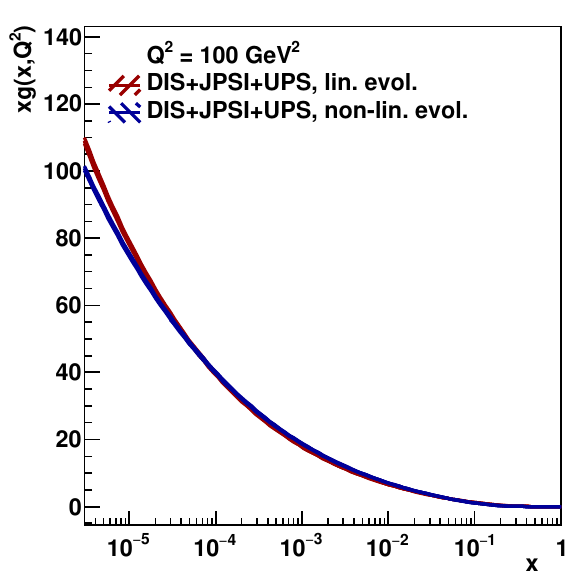}
\caption{\sf The gluon distributions at scales $Q^2=22~{\rm GeV}^2$ (left) and $100~{\rm GeV}^2$ (right) 
 obtained {\em with} (blue) and {\em without} (red) absorptive corrections, fitting DIS data combined with exclusive $J/\psi$ and $\Upsilon$ data. 
 The absorptive corrections were calculated for $R=3.6$ GeV$^{-1}$. 
}
\label{f24}
\end{center}
\end{figure*}

 Note that at  larger scales, for example $Q^2=22$ GeV$^2$, the low-$x$ gluon obtained with linear evolution exceeds that obtained with absorptive corrections by about 10\%, \textcolor{black}{reflecting the suppression of parton-density growth during the evolution by the absorptive corrections}, see Fig.~\ref{f24}. However, at $Q^2=m^2_Z$, the difference becomes almost negligible  $\sim 1-3$\% since \textcolor{black}{the dominant contribution to $xg(x,\mu_F^2)$ at large scales is generated through the $\mu_F^2$ evolution from moderately small $x$ values.}
  
\begin{figure*}[t]
\begin{center}
\hspace{-.5cm}
\includegraphics[scale=0.8]{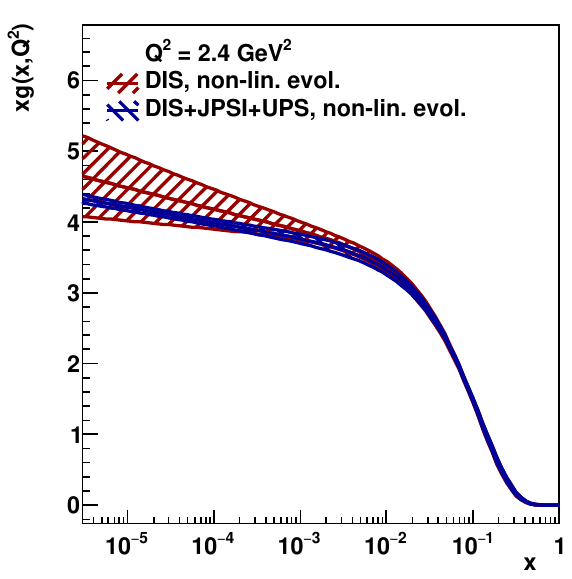}
\qquad
\includegraphics[scale=0.8]{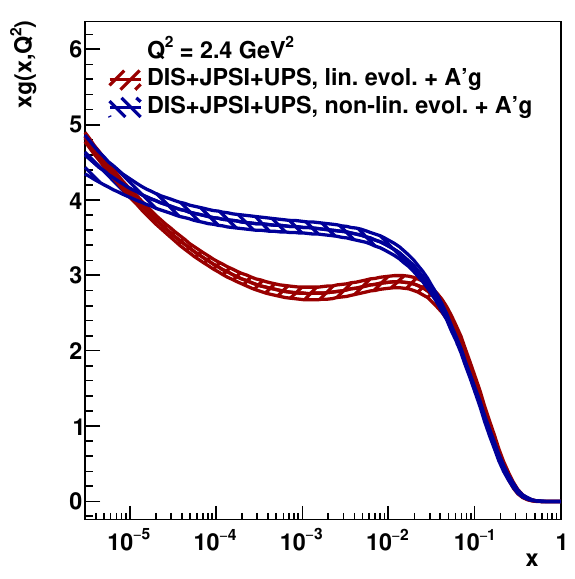}
\caption{\sf (Left:) The gluon distributions at scale $Q^2=2.4$ GeV$^2$
 obtained {\em with} absorptive corrections, fitting DIS data (red) and DIS data combined with exclusive $J/\psi$ and $\Upsilon$ data (blue). (Right:) The gluon distributions at scale $Q^2=2.4$ GeV$^2$
 obtained {\em with} (blue) and {\em without} (red) absorptive corrections, fitting DIS data combined with exclusive $J/\psi$ and $\Upsilon$ data. Here, the `negative gluon component' term proportional to $A'_{ g}$ is now kept in the input gluon PDF parametrisation, see text for details. The absorptive corrections were calculated for $R = 3.1$ GeV$^{-1}$ for the fit with just the DIS data and $R = 3.6$ GeV$^{-1}$ for the combined fit.  
}
\label{f23}
\end{center}
\end{figure*}

 In Fig.~\ref{f23} (left) we compare the input ($Q^2=2.4$ GeV$^2$) gluons coming from the fit of the only DIS data (red) with that from the fit where the DIS data were supplemented by the exclusive $J/\psi$ and $\Upsilon$ data.
 Both fits account for non-linear corrections. Fig.~\ref{f23} (right) shows the results of the fit where, instead of the non-linear corrections, the negative gluon component with $A'_{g}\neq 0$ is added to the input, see eqn.~(\ref{secondt}). \textcolor{black}{Table~\ref{tab:4} of Appendix A summarises the partial $\chi^2$ per degree of freedom for each dataset considered.} \textcolor{black}{Comparing the right columns of Tables 2 and 4, the description of the DIS data remains essentially unchanged, and thus these data can be described without the $A'_g$ term. On the other hand, the inclusion of this term leads to a very good description of the $J/\psi$ data, with $\chi^2/{\rm d.o.f} \ll 1$, reflecting the limited overlap in $x$ between the DIS and $J/\psi$ datasets. This leaves the $A'_g$ term relatively unconstrained by the DIS data and allows it to accommodate the $J/\psi$ data particularly well.} 

As it is seen in Fig.~\ref{f23} for $x>0.001$, the result of the fit without
the absorptive corrections but with the $A'_{g}$ term included  is rather close to that shown by red in Fig.~\ref{f21} (left) where only the DIS data were described. The $A'_{g}$ term is used just to fit the very low-$x$ vector-meson data ($g_{{\rm eff}}(x)$ points coming from exclusive $J/\psi$ production; there is no DIS data at so small $x$). The low input gluon density in the interval $10^{-4}<x<3 \times 10^{-2}$ now mimics the absorptive effects at a larger $Q^2$.
  
 Note that even having a larger number of free parameters (second term of eqn.~\eqref{secondt}) the value of the $\chi^2$ in the fit without the absorptive effects turns out to be larger (by about a factor of 20) than that in the fit with   
absorptive corrections included.

\begin{figure*}[t]
\begin{center}
\hspace{-.5cm}
\includegraphics[scale=0.75]{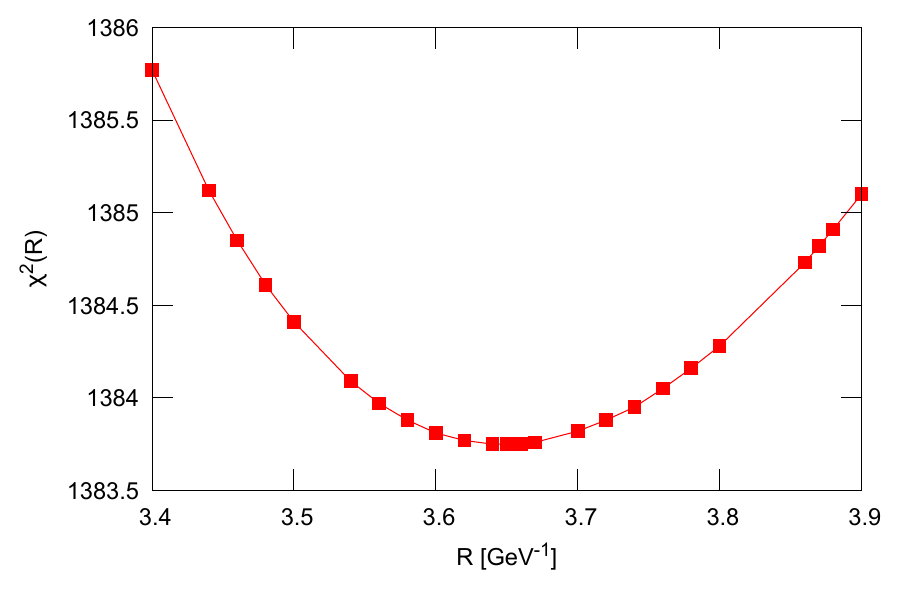}
\hspace{.5cm}
\caption{\sf Dependence of $\chi^2(R)$ on the strength of the non-linear correction, as determined by the parameter $R$. Here, the `negative gluon component' term proportional to $A'_{g}$ is fixed to zero. 
}
\label{f13}
\end{center}
\end{figure*}

\begin{figure*}[t]
\begin{center}
\hspace{-.5cm}
\includegraphics[scale=0.8]{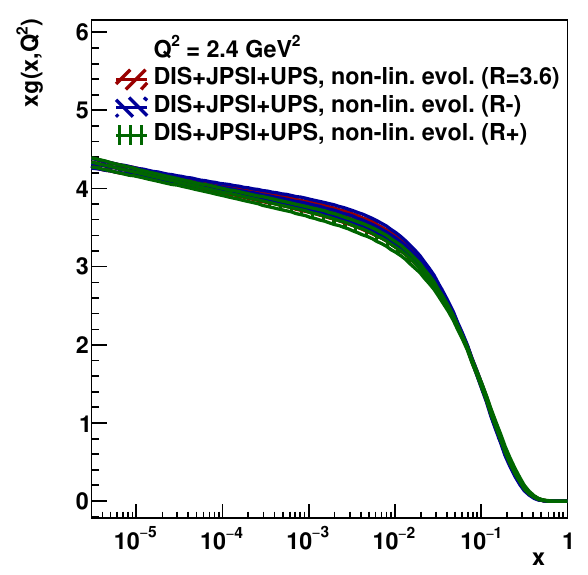}
\qquad
\includegraphics[scale=0.8]{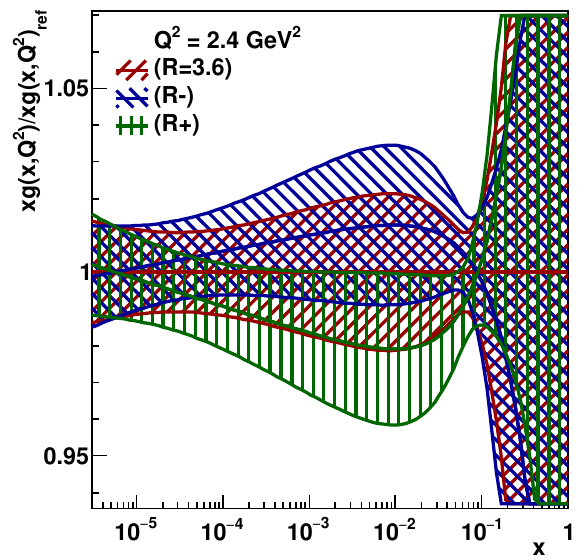}
\caption{\sf (Left:) The gluon distributions at scale $Q^2=2.4$ GeV$^2$ obtained {\it with} absorptive corrections for $R=\bar{R}=3.64$ GeV$^{-1}$~(\mbox{red}), corresponding to $\chi^2=\chi^2_{\min}$, and $R=R^+=3.87~(\mbox{green})$ and $R=R^-=3.46~(\mbox{blue})$ GeV$^{-1}$, corresponding to $\chi^2=\chi^2_{\min}+1$. Here, the `negative gluon component' term proportional to $A'_g$ is fixed to zero. \textcolor{black}{(Right:) The gluon PDFs obtained with $R=R^+$ and $R=R^-$ normalised to that obtained with $R=\bar{R}$.}
}
\label{f25}
\end{center}
\end{figure*}

We remark that the error bands shown in Figs.~\ref{f21},~\ref{f24}, and~\ref{f23} do not include the uncertainty associated with the parameter $R$. This contribution is shown separately in Fig.~\ref{f25} and is obtained by considering the standard 1-sigma tolerance, $\Delta \chi^2 = 1$, around the minimum value $\chi^2_\text{min}$ at the best-fit value $R=\bar{R}$. The dependence of the fit quality $\chi^2$ on $R$ is shown in Fig.~\ref{f13}, where a clear minimum is observed at $R\simeq 3.6$~GeV$^{-1}$. The resulting spread in the  gluon PDF is shown in Fig.~\ref{f25}. 

 The value of $R$ obtained in our fit can be considered as the measure of the ``hot-spot'' size. Indeed, while $R=R_N=5$~GeV$^{-1}$~=~1 fm is the proton size, the obtained $R=3.6$~GeV$^{-1}$ can be treated as the hot-spot radius ($R_{h.s.}$) for the case of {\em one} hot spot in the proton wave function.
 
 If there are $n$ hot spots, then
 \be
 \frac 1{R^2}=\frac{n-1}{nR^2_N}+\frac 1{nR^2_{h.s.}}\ .
 \label{Rhs}
 \ee
 That is 
 \be
 R^2_{h.s.}=\frac 1{n/R^2-(n-1)/R^2_N}\ .
 \ee
 Starting from $R=3.6$~GeV$^{-1}$ we obtain, with $n=3$, the value
 $R_{h.s.}=2.57$~GeV$^{-1}$.  In other words, from the present analysis we can for the first time evaluate the size of hot spots in the proton wave function. This was not done in~\cite{Gu}, since the $\chi^2$ dependence was found to be practically flat (see Fig.~3 therein).

\subsection{The NNLO* method}\label{subsec:NNLOstar}

\textcolor{black}{As proposed in Ref.~\cite{geff}, the method of ``effective gluon points'' used above can be generalised to NNLO. Although the coefficient functions for $J/\psi$ and $\Upsilon$ photo- and electro-production are currently known only at NLO~\cite{ISSK, SPJ, FGJT}, the effective gluon points may still be included in a NNLO global analysis through an approximate treatment. This is achieved by introducing a NNLO/NLO $K$-factor, whose value is inferred from the description of the available HERA $\gamma p \rightarrow J/\psi p$ (and/or $\gamma p \rightarrow \Upsilon p$) data.
}

The reason is as follows. 
In general, the NNLO/NLO $K$-factor depends on both the factorisation scale $\mu_F$ and the variable
$z=2\xi/(X+\xi)$, which is related to the ratio of the parton momentum fraction $x=X+\xi$ to $\xi$.\textcolor{black}{\footnote{Here, $\xi$ is the longitudinal momentum asymmetry (``skewness'') between the incoming and outgoing hadrons.}} That is, formally, $K = K(X,\xi,\mu_F)$. Consequently, the $K$-factor should be included inside the convolution integral that defines the $\gamma p \rightarrow J/\psi p$ amplitude,
\begin{equation}
\label{conv}
{\cal M}(\gamma+p\to J/\psi+p)=
\sum_{i=g,q}N_i
\int_{-1}^{1}\frac{dX}{X}
F_i(X,\xi,\mu_F,t)
K(X,\xi,\mu_F)
C_i^{\rm NLO}(\xi/X,\mu_F,\mu_R),
\end{equation}
where the constants $N_i$ provide the correct normalisation, see e.g.~\cite{FMRT2}, while $F_i$ and $C_i$ denote GPD and the corresponding coefficient function, respectively.  In the present analysis, the factorisation and renormalisation scales are fixed at $\mu_F = \mu_R =  M_{J/\psi}/2$. The momentum transfer $t=t_{\rm min} = 4\xi^2 m_p^2/(1-\xi^2) \sim 0$, owing to the very small values of $\xi$, so we refrain from discussing the $t$-dependence of GPDs. In the following, we therefore suppress the dependence of the convolution integral on $t,\mu_F$ and $\mu_R$.

The convolution integral in eqn.~(\ref{conv}) can be written schematically as
\begin{equation}
\label{conv1}
{\cal M}(\gamma+p\to J/\psi+p) \sim
\sum_{i=g,q}
\int_{-1}^{1}dX~
w_i(X,\xi) ~K(X,\xi),
\end{equation}
where $w(X,\xi) = F_i(X,\xi) C_i^{\rm NLO}(\xi/X) $ is a `weight' function determined by the coefficient functions and the GPDs.

Suppose we define a normalised weight function
\begin{equation}
\label{conv2}
\hat{w}(X,\xi) = \frac{w(X,\xi)}{\int_{-1}^1 dX~ w(X,\xi)},
\end{equation}
then the amplitude can be written 
\begin{equation}
\label{conv3}
{\cal M}(\gamma+p\to J/\psi+p) \sim
\sum_{i=g,q} \left(
\int_{-1}^{1}dX~
w_i(X,\xi)\right)  \left(
\int_{-1}^{1}dX~
\hat{w}_i(X,\xi)~K(X,\xi)\right),
\end{equation}
where the second factor is the {\it weighted average} of the $K$-factor,
\begin{equation}
    \langle K \rangle = \int_{-1}^1 dX~ \hat{w}(X,\xi)  K(X,\xi)
\end{equation}

If this weighted average is approximately independent of $\xi$, then we may write $\langle K \rangle = K_{\rm eff} = {\rm const}.$ Indeed, the coefficient function is dominated by the region $X\sim\xi$, corresponding to $z=\mathcal O(1)$, see~\cite{Flett:2019pux}. The coefficient function does not introduce a power enhancement at small $x$ and, at NNLO, additional logarithms from higher-order gluon-loop insertions behave like $\ln^n(z)$, which grow more slowly than any power. The $z$ dependence of the $K$-factor is therefore averaged over with a weight determined by the gluon distribution. If the latter exhibits an approximately scale-independent power behaviour $xg(x) \propto x^{-\lambda}$, the shape of this weight remains approximately unchanged over the relevant $X$ range, allowing the convolution of $K(z)$ to be represented by an approximately constant effective NNLO/NLO $K$-factor.\footnote{The weight function changes with $\xi$ as the gluon PDF of functional form $xg(x) \propto x^{-\lambda}$ evolves. If $\lambda$ is approximately constant, changing $\xi$ only rescales the overall normalisation, $g(\xi) \sim \xi^{-\lambda}$,
but it does not strongly change the shape of the distribution inside the integral.
Therefore the relative sampling of $z$ values remains approximately the same, meaning
$$
\frac{\int dX~w(X,\xi) K(X,\xi)}{
\int dX~w(X,\xi)} \simeq {\rm const.}
$$
This constant is the effective NNLO/NLO $K$-factor.} As can be seen from Fig.~\ref{f25} (left), the obtained low-$x$ gluon behaviour at our optimal scale~\cite{opt,opt2,Q0} $\mu_F = M_{J/\psi}/2$ is consistent with $\lambda = {\rm const}$.

The value of the $K$-factor can therefore be determined by describing exclusive $J/\psi$ photoproduction data from HERA using NNLO partons together with the NLO coefficient functions in the region
$0.01 > x > 0.001$, where the uncertainty of the gluon distribution from present global analyses is relatively small. Specifically, the square of the $K$-factor is obtained from the ratio of the measured $\gamma + p \to J/\psi + p$ cross section, $\sigma$(data), to the cross section $\sigma^{\mbox{NLO/NNLO}}$ calculated
using the NLO coefficient functions but with NNLO input partons:
\begin{equation}
\label{e12}
K=\sqrt{\frac{\sigma(\mbox{data})}{\sigma^{\mbox{NLO/NNLO}}}}~.
\end{equation}
The final value of $K$ is then obtained by taking the average of the values extracted from the individual data points.

\begin{figure*}[t]
\begin{center}
\hspace{-.5cm}
\includegraphics[scale=0.75]{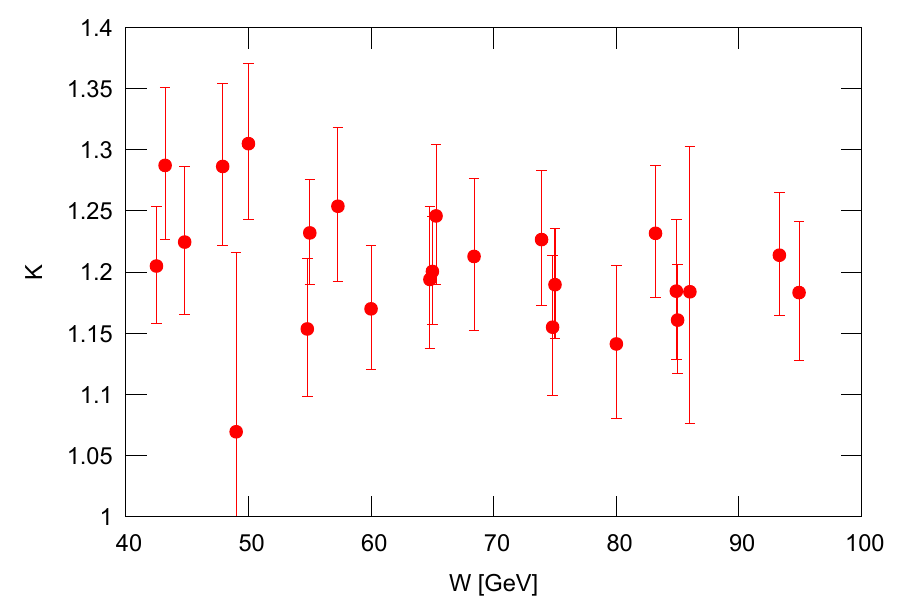}
\vspace{0.5cm}
\caption{\sf Dependence of the $K$-factor value on the photon-proton energy,$\sqrt s=W$.
}
\label{f14}
\end{center}
\end{figure*}

Indeed, as it is seen in Fig.~\ref{f14}, within its uncertainties the $K$-factor may be considered as the constant 
\begin{equation}
K=1.2\pm 0.1.
\end{equation}
 
This $K$ factor was included into \texttt{xFitter} by 
rescaling the effective gluon points $g_{{\rm eff}}(x)$. The uncertainty associated with this renormalisation was treated as a correlated systematic uncertainty among all effective gluon points. In practice, the rescaled points are multiplied by a common nuisance parameter $K'$, which allows for a correlated shift of the entire dataset, with the constraint $1/K'=1.2\pm 0.1$. Thus, the uncertainty on the $K$-factor is propagated consistently through the fit as a correlated systematic error rather than as independent uncertainties on the individual $g_{\rm eff}(x)$ points. These points were then used in a parton analysis within \texttt{xFitter} at NNLO accuracy. We dub such an analysis the NNLO* method, to distinguish it from an analysis in which the full NNLO coefficient functions would be used, when they become available. \textcolor{black}{In such an analysis, the DIS structure functions meanwhile are treated fully at NNLO in the FONLL scheme.}

\begin{figure*}[t]
\begin{center}
\hspace{-.5cm}
\includegraphics[scale=0.8]{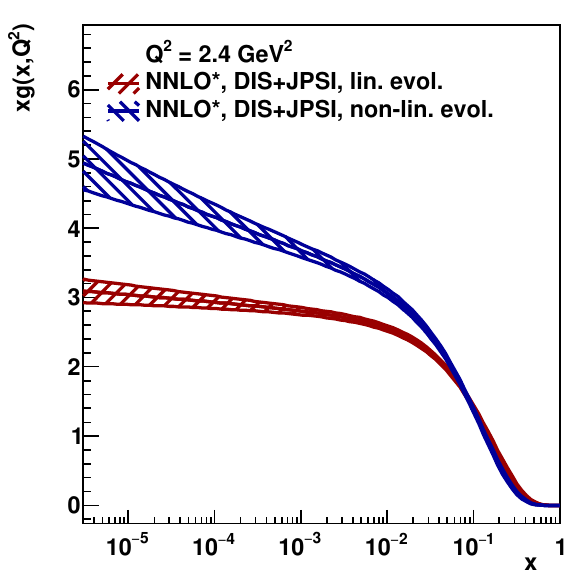}
\qquad
\includegraphics[scale=0.8]{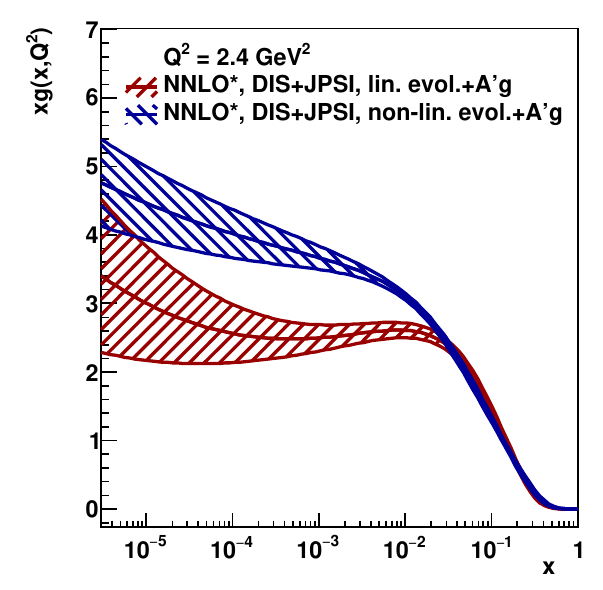}
\caption{\sf The gluon distributions at scale $Q^2=2.4$ GeV$^2$  obtained in the NNLO$^*$ approach {\em with} (blue) and {\em without} (red) absorptive corrections. In the right panel, the `negative gluon component' term proportional to $A'_g$ is kept in the input gluon PDF parametrisation, see text for details.  The absorptive corrections in the left panel were calculated for $R=3.27$ GeV$^{-1}$, while in the right with $R=3.23$ GeV$^{-1}$, corresponding to $R=\bar{R}$ in each case. 
}
\label{f15}
\end{center}
\end{figure*}

\begin{figure*}[t]
\begin{center}
\hspace{-.5cm}
\includegraphics[scale=0.75]{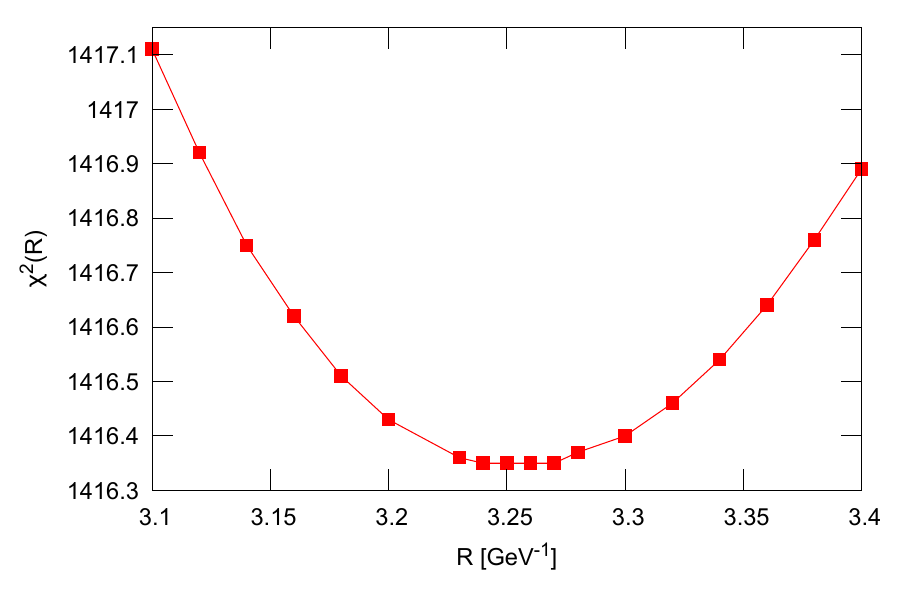}
\caption{\sf Dependence of $\chi^2(R)$ on the strength of the non-linear correction, as determined by the parameter $R$, in the NNLO* method. Here, the `negative gluon component' term proportional to $A'_{g}$ is fixed to zero. 
}
\label{f16}
\end{center}
\end{figure*}

\textcolor{black}{
The results are shown in Figs.~\ref{f15} and \ref{f16}. Both with and without the inclusion of the $A'_{g}$ term in eqn.~\eqref{secondt}, the fits incorporating the non-linear corrections provide a significantly better description of the data. For instance, without the $A'_{g}$ contribution, the fit including non-linear effects yields $\chi^2=1416$ for 1166 degrees of freedom, compared to $\chi^2=1495$ when non-linear effects are neglected.
Furthermore, once the absorptive (non-linear) effects are explicitly included, the additional $A'_{g}$ term leads to only a marginal improvement in the fit quality, reducing the total $\chi^2$ from 1416 to 1412. This suggests that the $A'_{g}$ contribution in fits without non-linear corrections effectively parametrises effects that are otherwise generated by absorptive corrections. Tables~\ref{tab:5} and~\ref{tab:6} of Appendix A summarise the partial $\chi^2$ per degree of freedom for each dataset considered.}

\textcolor{black}{In the case of the NNLO$^*$ fit, we obtain a smaller best-fit value $R \sim 3.3$ GeV$^{-1}$ as compared to the NLO description, where $R \sim 3.6$ GeV$^{-1}$ was obtained.  This means that the smaller value of the NNLO gluon PDF from global analyses (in comparison with that
at NLO) is compensated by a smaller $R$ to have the same non-linear
effect. }

 \section{Conclusions}
We have described the implementation of non-linear (absorptive) effects~\cite{GLR,MQ} into the DGLAP evolution generated by 
 \texttt{APFEL++} \cite{Bertone:2013vaa,Bertone:2017gds}, and have used \texttt{xFitter}~\cite{xF} to perform fits to DIS data from HERA and effective exclusive heavy-vector meson production data from the LHC, taking these non-linear effects into account. Heavy-vector meson production data were described using
the  optimal scale $\mu_F=M_V/2$,
which effectively resums the double-logarithmic family of terms $(\alpha_s \ln(Q^2)\ln(1/\xi))^n$ into the input PDFs, resulting in improved scale
stability~\cite{opt,opt2,Q0}.

\textcolor{black}{Our results indicate that accounting for these corrections improves the $\chi^2$ quality of the fit, reduces PDF uncertainties, and avoids the spurious unphysical local minimum in the gluon distribution at $\mu_F^2 \simeq 2.4$ GeV$^2$ and $x \sim 10^{-3}$, encountered in some global PDF analyses. This suggests that absorptive effects should be considered in future such analyses aiming to constrain the gluon density in the small-$x$, low-$Q^2$ region.}

Based on the fit \textcolor{black}{at NLO} we evaluate, for the first time, the ``hot-spot'' size $R_{h.s}$ to be $R_{h.s}\simeq 2.5 - 2.6$~GeV$^{-1}$, about twice smaller than the proton size. However, the hot spot size is still rather large. Therefore, the effect of saturation is hardly seen in the present data.

Analyses at both NLO and approximate NNLO have been considered. Since the coefficient function for heavy-vector meson photoproduction is not yet calculated at NNLO accuracy,  
we carried out the so-called NNLO$^*$ method, introduced in our previous~\cite{geff}. Here, the NNLO coefficient function $C^{{\rm NNLO}}$ is replaced by $KC^{{\rm NLO}}$ with the constant NNLO/NLO $K$-factor extracted from a fit of exclusive $J/\psi$ photoproduction data from HERA.

Once non-linear evolution is included, the `negative gluon component' that is proportional to $A'_g$, is effectively washed out, as can be seen by comparing Fig.~6 (NLO) and Fig.~10 (NNLO*). In other words, when this term is included in fits based on non-linear evolution, it becomes essentially redundant: the resulting gluon PDF is almost unchanged compared with a fit using non-linear evolution alone.\footnote{A similar observation was made in the analysis of~\cite{xFitterDevelopersTeam:2018hym}, where the $A'_g$ term was found to be unnecessary once $\ln(1/x)$ resummation was included.} This contrasts with the case of linear evolution, where the inclusion of the negative gluon component has a greater impact. In particular, adding this term to the linear-evolution fit drives the gluon PDF upwards too strongly at very small $x$ for $Q^2=2.4$~GeV$^2$.

\textcolor{black}{The present study therefore demonstrates that absorptive corrections can be consistently incorporated into a modern PDF fitting framework, providing an improved description of both inclusive DIS and exclusive heavy vector meson production data within a combined analysis. The implementation of the non-linear evolution within \texttt{APFEL++} will be made publicly available as a separate module through \texttt{xFitter}, together with the corresponding PDF sets in LHAPDF format.}

 \section*{Appendix A: Tables of $\chi^2$}

\begin{table}[ht]
\centering
\begin{tabular}[t]{lcccc}
\toprule
Dataset&$\chi_{\text{min}}^2/\text{d.o.f}$~(DIS, lin. evol.) & $\chi_{\text{min}}^2/\text{d.o.f}$~(DIS, non-lin. evol.)&\\
\midrule
HERA1+2 NCep 820&85/73&81/73\\
HERA1+2 NCep 460&224/207&226/207\\
HERA1+2 CCep&43/39&44/39\\
HERA1+2 NCem&218/159&222/159\\
HERA1+2 CCem&53/42&52/42\\
HERA1+2 NCep 575&221/257&222/257\\
HERA1+2 NCep 920&456/391&446/391\\
\midrule
Total $\chi_{\text{min}}^2/\text{d.o.f}$&1389/1156 $\approx$ 1.20& 1367/1156 $\approx$ 1.18\\
\bottomrule
\end{tabular}
\caption{\sf{The partial $\chi_{\text{min}}^2/\text{d.o.f}$ for each dataset included in the fit of DIS data using linear DGLAP evolution (left) and non-linear DGLAP evolution (right). Here, we use the DIS data in the kinematic range $Q^2 > 2.4$ GeV$^2$. The total $\chi_{\text{min}}^2/\text{d.o.f}$ is also given.}} 
\label{tab:2}
\end{table}

\begin{table}[ht]
\centering
\begin{tabular}[t]{lcccc}
\toprule
Dataset&$\chi_{\text{min}}^2/\text{d.o.f}$~(DIS+JPSI+UPS, & $\chi_{\text{min}}^2/\text{d.o.f}$~(DIS+JPSI+UPS,&\\
& lin. evol.) & non-lin. evol.)&\\
\midrule
HERA1+2 NCep 820&82/73&82/73\\
HERA1+2 NCep 460&225/207&227/207\\
HERA1+2 CCep&49/39&44/39\\
HERA1+2 NCem&226/159&221/159\\
HERA1+2 CCem&52/42&52/42\\
HERA1+2 NCep 575&233/257&222/257\\
HERA1+2 NCep 920&491/391&450/391\\
LHC excl. $J/\psi$ $pp$ 13 TeV&7.1/10& 12.6/10\\
LHC excl. $\Upsilon$ $pp$ 7,8 TeV&4.1/3& 2.9/3\\
\midrule
Total $\chi_{\text{min}}^2/\text{d.o.f}$&1464/1169 $\approx$ 1.25& 1384/1169 $\approx$ 1.18\\
\bottomrule
\end{tabular}
\caption{\sf{The partial $\chi_{\text{min}}^2/\text{d.o.f}$ for each dataset included in the fit of DIS and exclusive $J/\psi, \Upsilon$ data using linear DGLAP evolution (left) and non-linear DGLAP evolution (right). Here, we use the DIS data in the kinematic range $Q^2 > 2.4$ GeV$^2$. The total $\chi_{\text{min}}^2/\text{d.o.f}$ is also given.}} 
\label{tab:3}
\end{table}

\begin{table}[ht]
\centering
\begin{tabular}[t]{lcccc}
\toprule
Dataset&$\chi_{\text{min}}^2/\text{d.o.f}$~(DIS+JPSI+UPS, & $\chi_{\text{min}}^2/\text{d.o.f}$~(DIS+JPSI+UPS,&\\
& lin. evol. and term $\propto A'_g$) & non-lin. evol. and term $\propto A'_g$)&\\
\midrule
HERA1+2 NCep 820&82/73&83/73\\
HERA1+2 NCep 460&222/207&227/207\\
HERA1+2 CCep&44/39&43/39\\
HERA1+2 NCem&215/159&221/159\\
HERA1+2 CCem&52/42&52/42\\
HERA1+2 NCep 575&222/257&221/257\\
HERA1+2 NCep 920&455/391&449/391\\
LHC excl. $J/\psi$ $pp$ 13 TeV&10.5/10& 2.8/10\\
LHC excl. $\Upsilon$ $pp$ 7,8 TeV&2.7/3& 2.5/3\\
\midrule
Total $\chi_{\text{min}}^2/\text{d.o.f}$&1402/1167 $\approx$ 1.20& 1375/1167 $\approx$ 1.18\\
\bottomrule
\end{tabular}
\caption{\sf{The partial $\chi_{\text{min}}^2/\text{d.o.f}$ for each dataset included in the fit of DIS and exclusive $J/\psi, \Upsilon$ data using linear DGLAP evolution (left) and non-linear DGLAP evolution (right) including the `negative gluon component' term proportional to $A'_g$, see text for details. Here, we use the DIS data in the kinematic range $Q^2 > 2.4$ GeV$^2$. The total $\chi_{\text{min}}^2/\text{d.o.f}$ is also given.}} 
\label{tab:4}
\end{table}

\begin{table}[ht]
\centering
\begin{tabular}[t]{lcccc}
\toprule
Dataset&$\chi_{\text{min}}^2/\text{d.o.f}$~(NNLO*,& $\chi_{\text{min}}^2/\text{d.o.f}$~(NNLO*,&\\
&DIS+JPSI, lin. evol.) &DIS+JPSI, non-lin. evol.)&\\
\midrule
HERA1+2 NCep 820&79/73&77/73\\
HERA1+2 NCep 460&219/207&224/207\\
HERA1+2 CCep&47/39&45/39\\
HERA1+2 NCem&216/159&215/159\\
HERA1+2 CCem&54/42&55/42\\
HERA1+2 NCep 575&233/257&226/257\\
HERA1+2 NCep 920&496/391&453/391\\
LHC excl. $J/\psi$ $pp$ 13 TeV&17.8/10& 4.7/10\\
\midrule
Total $\chi_{\text{min}}^2/\text{d.o.f}$&1495/1166 $\approx$ 1.28& 1416/1166 $\approx$ 1.21\\
\bottomrule
\end{tabular}
\caption{\sf{The partial $\chi_{\text{min}}^2/\text{d.o.f}$ for each dataset included in the fit of DIS and exclusive $J/\psi, \Upsilon$ data using linear DGLAP evolution (left) and non-linear DGLAP evolution (right) in the NNLO* approach, see text for details. Here, we use the DIS data in the kinematic range $Q^2 > 2.4$ GeV$^2$. The total $\chi_{\text{min}}^2/\text{d.o.f}$ is also given.} } 
\label{tab:5}
\end{table}

\begin{table}[ht]
\centering
\begin{tabular}[t]{lcccc}
\toprule
Dataset&$\chi_{\text{min}}^2/\text{d.o.f}$~(NNLO*, DIS+JPSI,& $\chi_{\text{min}}^2/\text{d.o.f}$~(NNLO*, DIS+JPSI,&\\
&lin. evol. and term $\propto A'_g$) &non-lin. evol. and term $\propto A'_g$)&\\
\midrule
HERA1+2 NCep 820&79/73&77/73\\
HERA1+2 NCep 460&220/207&225/207\\
HERA1+2 CCep&44/39&45/39\\
HERA1+2 NCem&214/159&218/159\\
HERA1+2 CCem&57/42&54/42\\
HERA1+2 NCep 575&228/257&226/257\\
HERA1+2 NCep 920&474/391&458/391\\
LHC excl. $J/\psi$ $pp$ 13 TeV&3.5/10& 4.4/10\\
\midrule
Total $\chi_{\text{min}}^2/\text{d.o.f}$&1455/1164 $\approx$ 1.25& 1412/1164 $\approx$ 1.21\\
\bottomrule
\end{tabular}
\caption{\sf{The partial $\chi_{\text{min}}^2/\text{d.o.f}$ for each dataset included in the fit of DIS and exclusive $J/\psi, \Upsilon$ data using linear DGLAP evolution (left) and non-linear DGLAP evolution (right) in the NNLO* approach and including the `negative
gluon component' term proportional to $A'_g
$, see text for details. Here, we use the DIS data in the kinematic range $Q^2 > 2.4$ GeV$^2$. The total $\chi_{\text{min}}^2/\text{d.o.f}$ is also given.} } 
\label{tab:6}
\end{table}

  \section* {Acknowledgments}  
C.~A.~F. is supported by the European Union’s Horizon Europe research and innovation programme under the Marie Skłodowska-Curie grant agreement No. 101204057 [AutomOnium]. The work of V.B. has
been supported by l’Agence Nationale de la Recherche (ANR), project ANR-24-CE31-7061-01.

\end{document}